# Tuning of the Quantum-Confined Stark Effect in Wurtzite [000$\bar{1}$] Group-III-Nitride Nanostructures by the Internal-Field-Guarded-Active-Region Design

S. Schlichting[1,*], G. M. O. Hönig[2], J. Müßener[3], P. Hille[3], T. Grieb[3], J. Teubert[4], J. Schörmann[4], M. R. Wagner[1], A. Rosenauer[3], M. Eickhoff[3], A. Hoffmann[1], and G. Callsen[1,†]

*1: Institut für Festkörperphysik, Technische Universität Berlin, Hardenbergstr. 36, 10623 Berlin, Germany*
*2: Bundesanstalt für Materialforschung und -prüfung (BAM), 12200 Berlin, Germany*
*3: Institut für Festkörperphysik, Universität Bremen, Otto-Hahn-Allee 1, 28359 Bremen, Germany*
*4: I. Physikalisches Institut, Justus-Liebig Universität Giessen, Heinrich-Buff-Ring 16, 35392 Giessen, Germany*

**Recently, we suggested an unconventional approach [the so-called Internal-Field-Guarded-Active-Region Design (IFGARD)] for the elimination of the crystal polarization field induced quantum confined Stark effect (QCSE) in polar semiconductor heterostructures. And in this work, we demonstrate by means of micro-photoluminescence techniques the successful tuning as well as the elimination of the QCSE in strongly polar [000$\bar{1}$] wurtzite GaN/AlN nanodiscs while reducing the exciton life times by more than two orders of magnitude. The IFGARD based elimination of the QCSE is independent of any specific crystal growth procedures. Furthermore, the cone-shaped geometry of the utilized nanowires (which embeds the investigated IFGARD nanodiscs) facilitates the experimental differentiation between quantum confinement- and QCSE-induced emission energy shifts. Due to the IFGARD, both effects become *independently* adaptable.**

## I. INTRODUCTION

Group-III-nitride semiconductors are key materials for visible and ultraviolet LEDs, LDs,[1–3] and quantum-light sources.[4–6] In particular GaN as well as AlN favorably crystalize in the wurtzite crystal structure.[7] Hence, heterostructures based on these materials suffer from a strong electric field induced by a piezo- and pyroelectric polarization parallel to the most natural crystal growth direction [0001], the so called *c* axis.[8–12] The polarization-induced internal fields cause a redshift of the exciton emission energy inside these heterostructures, known to be the prominent feature of the quantum-confined Stark effect (QCSE),[13–17] which is accompanied by a drastic decrease of the spatial electron-hole overlap in the direction of the *c* axis.[4,9,18–27] Different approaches to eliminate or to diminish the electric field in group-III-nitride heterostructures (across the optically active region) have been investigated, such as growth on non- or semi-polar crystal planes,[28,29] forcing the growth of the cubic zincblende phase,[30] or by screening the fields with doping-induced free carriers.[31] Generally, approaches to avoid the preferential [0001] wurtzite crystal growth are challenging, slow, and they often produce a reduced crystal quality.[32–37] A more promising method to control the internal electric field is the Internal-Field-Guarded-Active-Region Design (IFGARD),[38,39] theoretically developed by Hönig et al..[40] As described in detail in Ref. 40, a conventional structure of a GaN quantum well (QW) embedded in AlN barriers is complemented in the IFGARD structure by additional GaN guard layers enclosing the AlN barriers [see Fig. 1(a) + (c)]. This is not intuitive as the additional GaN guard layers reabsorb a particular percentage of the photons generated in the QW. But as discussed by Hönig et al.,[40] the overall gain based on the elimination of the polarization field in the QW and the resulting boost in the exciton recombination probability, can overcompensate the reabsorption losses by the guard layers, if the thickness of the guard layer in the emission direction is below the emitted wavelength.[41–44] It was numerically demonstrated that this concept leads to a constant piezo- and pyroelectric polarization potential inside of a QW, which results in flat conduction- and valence-band edges therein [Fig. 1 (c)]. This elimination of the electric field inside the QW results in a higher spatial overlap of the electron and hole wave function, and hence in a drastic rise of the exciton recombination probability. As calculated in Ref. 40 this maximized exciton recombination probability implicates a gain in the emitted single photons by 2 orders of magnitude.

---

[*] sarah.schlichting@tu-berlin.de
[†] Present address: Institute of Physics, École Polytechnique Fédérale de Lausanne (EPFL), CH-1015 Lausanne, Switzerland

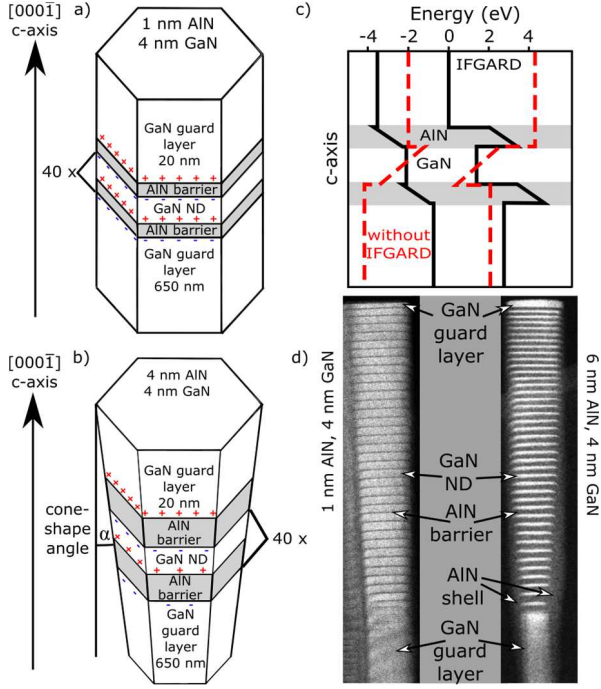

**FIG. 1.** Schematic sketch of nanowires (NWs) with embedded IFGARD nanodiscs with AlN barriers of 1 nm (a) and 4 nm (b) thickness. NWs with an increased barrier thickness are accompanied by an increased cone-shape angle α due to lateral growth of the barrier material. Red + and blue - mark the positions of interface charges induced by the wurtzite crystal polarization. Due to the cone-shape angle α of the NWs (b), the individual net interface charges of each AlN-barrier interface differ. In (c), a sketch of the black IFGARD band profile is compared to the corresponding conventional nanodisc band profile in red. (d) shows HAADF-STEM images of the stacked GaN/AlN nanodiscs with 1 nm (left) and 6 nm barriers (right).

It is important to note that the IFGARD approach does not require a change in the underlying growth procedure; hence, no degradation of the crystal quality occurs as frequently observed for non- and semi-polar growth. Maintaining the initial crystal growth method and substrate material is a big advantage of the present concept compared to any other alternative approaches that aim to reduce or even eliminate the internal polarization fields in semiconductor heterostructures.[32–37]

In this work, we show the successful implementation of the IFGARD concept for the example of nanowires (NWs) containing GaN nanodiscs (NDs). In particular, we demonstrate the reduction of the internal electric field inside of the GaN NDs down to zero, which reveals the immense advantage of the IFGARD over alternative, more conservative design concepts. Furthermore, we identify the individual contributions of the pure QCSE and the pure confinement effect based on two sample series with varying barrier and ND thickness.

## II. SAMPLES

The investigated NWs were grown by plasma-assisted molecular beam epitaxy ($T_{substrate}$ = 790 °C, $T_{Ga}$ = 916 °C, $T_{Al}$ = 1069 °C) on Si (111) substrates.[21,31] Figure 1 illustrates the GaN NWs [Fig. 1(a+b)] with the corresponding band profiles along the growth axis [Fig. 1(c)] as well as corresponding Scanning-Transmission-Electron-Microscopy (STEM) images [Fig. 1(d)]. The STEM images were recorded with a HAADF (High-Angle Annular Dark Field) detector showing single NWs with 1-nm-thick AlN barriers (left, black) and 6-nm-thick AlN barriers (right, black) at a constant ND (white) thickness of 4 nm.

The NWs grow along the polar $[000\bar{1}]$ axis of the wurtzite crystal structure. Each NW embeds a stack of 40 GaN NDs separated by AlN barriers. The first GaN guard layer has a length (in growth direction) of 650 nm. The NW ends with a 20-nm-thick GaN guard layer [Fig. 1(a+b)] in order to realize the fundamental IFGARD symmetry. It is important to underline that the most simplistic stack element consists of one AlN barrier followed by one GaN ND[40] being repeated 40-times in order to clearly separate the ND from the guard layer signal. This sequence is encapsulated by the aforementioned GaN guard layers. In total, the NWs have a length of up to 1 μm.

The formation of an AlN shell and an increasing diameter of successively grown GaN NDs is observed due to the significant lateral growth of AlN at the applied temperatures [Fig. 1(d)].[45] Consequently, the NWs exhibit a cone-shaped structure[46] with the diameter increasing from 28 to 36 nm for the NWs with 1-nm-thick AlN barriers [Fig. 1(d), left] and from 38 to 74 nm for the NWs with 6-nm-thick AlN barriers [Fig. 1(d), right]. Considering the conservation of a constant interface-charge density, the particular cone shape leads to different net interface charges [red +, blue - in Fig. 1(a+b)] at either side of each AlN barrier; because the top interface to the GaN is larger than the interface to the GaN below each AlN barrier. This total interface-charge discrepancy is the perfect feature to proof the influence of the IFGARD on the electric field within the NDs, as in this case the flatness of the band-edge [Fig. 1(c)] depends on the barrier thickness - similar to what Hönig et al. discussed in Ref. 40 for IFGARD-based quantum dots. Hence, it becomes possible to control the built-in electric field in the NDs by adjusting the AlN barrier thickness and thus the net interface charges. Furthermore, thicker AlN barriers also have an enhanced lateral dimension, which leads to a more pronounced cone-shape of such NWs with thicker AlN barriers, further enhancing their built-in electric field strengths. The angle α of the cone-shaped NW, delineated in Fig. 1(b), gets smaller with reduced AlN barrier thickness until the cone-shape turns into a straight shape [Fig. 1(a) vs. 1(b, d)]. Consequently, we produced two sample series: i) samples with decreasing AlN barrier thickness from 6 nm to 1 nm with a constant GaN ND thickness of 4 nm and ii) samples with a constant AlN barrier thickness of 1 nm and a GaN ND thicknesses of 1, 2, and 3 nm. Since a ND thickness of 4 nm is large compared to the exciton-Bohr radius in GaN (effective Bohr-radius $a^* = \frac{m_e}{m^*}\varepsilon_r a_0 \leq 3\,nm$),[4,47–51] we would expect an emission energy (without built-in electric fields) close to the free exciton transition energy in bulk GaN of approximately 3.48 eV at cryogenic temperatures.[52,53] Hence, decreasing the thickness of the AlN barriers is expected to blueshift the ND emission towards the free exciton transition energy in bulk material by reducing the internal electric fields in the NDs. After having counteracted the QCSE by the discussed IFGARD structure with minimal barrier widths, we tune the energy to even higher energies by confining the excitons in thinner NDs.

## III. EXPERIMENTAL DETAILS

The optical properties of the NDs were investigated by micro-photoluminescence (μ-PL) and time-resolved μ-PL (μ-TRPL) measurements, the latter with two different experimental setups to explore the exciton life times from the ps- up to the μs-range. The samples were mounted into a micro helium-flow cryostat providing measurement temperatures down to 5 K.
In the μ-PL setup the NWs were excited by a pulsed (76 MHz repetition rate, 5 ps pulse width and 2 μm spot size) and frequency-quadrupled fiber laser resulting in an excitation wavelength of 258 nm. The spectra were dispersed by a 0.85 m single monochromator by a 150 lines/mm grating (500 nm blaze). The photons were detected with an UV-enhanced Si-charge-coupled device (CCD) array. All spectra were calibrated with an Hg-lamp and are corrected for the refractive index of air. μ-TRPL measurements in the ps- to ns-range were performed with the same laser as used for the μ-PL measurements, here coupled into a subtractive 0.35-m double-monochromator with 2400 lines/mm gratings as dispersing elements (300 nm blaze). The time-correlated single-photon-counting technique is used in combination with a hybrid photomultiplier detector assembly for recording the temporal decays.
The μs-range TRPL measurements are performed with a pulsed dye laser (350 nm, 100 Hz, 20 ns pulse width) pumped by a 308 nm XeCL-Excimer laser (100 Hz, 20 ns pulse width). For the dispersion an additive double monochromator with a focal length of 1 m and a holographic 1800 lines/mm grating are used. For recording the PL transients a multichannel-plate photomultiplier with an S20 cathode combined with time-correlated single-photon-counting and multi-channel-scaling technique (including multi-stop capability) are utilized. The recorded temporal decays are all corrected with the specific response characteristics of the respective setup.[54]

## IV. EXMPERIMENTAL OBSERVATIONS

Figure 2 displays μ-PL spectra, which are recorded with an excitation power of 15 μW at a temperature of 7 K. 15 μW excitation-power spectra are chosen for this plot as they represent a trade-off between energetically well separated luminescence maxima for all samples with enough intensity to be plotted on a linear scale (compare the yellow bar in Fig. 3). The spectra are not normalized and therefore quantitatively comparable in absolute intensities. Please note that the excitation power of 15 µW does not represent the lowest value, which is marked by a green bar in Fig. 3. Besides the dominant ND emission signature, the luminescence band of the bulk-like GaN guard layers - close to the free exciton emission energy in bulk GaN 3.48 eV[52,53] is visible and marked in Fig. 2 by the dashed, blue line. As shown in Fig. 2, the emission peak energy of the GaN NDs in the QCSE regime (GaN ND thickness is constant at 4 nm with varied AlN-barrier thickness: 6 nm, 4 nm, 2 nm and 1 nm) shifts from 2.76 eV to 3.34 eV for decreased barrier thicknesses. This shift is accompanied by a strong increase in absolute luminescence intensity combined with a decrease in the full width at half maximum (FWHM) (750 meV to 75 meV) of the ND-luminescence peaks in Fig. 2 for decreasing barrier thicknesses.

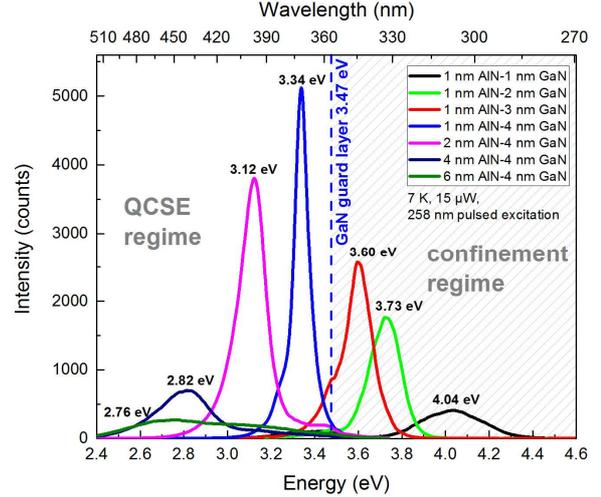

**FIG. 2.** The μ-PL spectra of all investigated samples are excited by a pulsed fiber laser with 258 nm wavelength and an excitation power of 15 μW at 7 K. In all samples the dominant GaN ND luminescence is accompanied by the GaN-guard-layer luminescence (marked by the dashed, blue line), which divides the samples series into the so called QCSE regime and the confinement regime.

For decreasing ND thicknesses an even further shift to higher emission energies from 3.34 eV to 4.04 eV occurs due to a stronger confinement of the charge carriers in the thinner NDs. At the same time, the absolute luminescence intensity decreases while the FWHM of the ND-luminescence peaks increases from 83 meV to 330 meV. The luminescence intensity decrease is caused by the reduced photon-absorption volume of the thinner NDs, while the stronger impact of monolayer-thickness variations in the thinner ND ensembles broadens the emission bands. As motivated at the beginning of this section, Fig. 2 was recorded with an increased excitation power of 15 µW causing a blue-shifted photoluminescence-peak position relative to the lowest pump powers in Fig. 3 due to partial screening of the QCSE by excited charge carriers.[55–57] The energetically smallest, measureable luminescence-peak positions are therefore displayed in Fig. 3. Figure 3 reveals the influence of the excitation-pump power that was varied from 0.035 µW to 500 µW on the emission-peak energy of the respective ND luminescence. On the right, vertical axis (blue) the emission energy shift is noted, representing the quantitative screening of the QCSE by the excited charge carriers plus a ND-band-filling effect converging for thinner NDs towards 12 meV. The grey shading visually divides Fig. 3 into a region either containing the results for IFGARD samples with a negligible QCSE (with total emission-energy shifts ≤ 14 meV) or those samples, whose emission characteristics are evidently affected by the QCSE. This grey shading is transferred to Fig. 2 including the notations: "QCSE" and "confinement regime". The energy range where the QCSE dominates the emission properties is located below the bulk-like GaN luminescence of the guard layers. As visible in Fig. 3, the total emission-energy shift reduces with decreasing AlN-barrier widths from 617 meV (6 nm AlN barrier) to 43 meV (1 nm AlN barrier). The apparent trends for the intensity, the energy shifts and the evolution of the FWHM values are accompanied by corresponding changes in the exciton lifetimes. The results of the µ-TRPL measurements are presented for all samples within the QCSE regime in Fig. 4 (a) and for the samples within the confinement regime in Fig. 4(b). As the transients exhibit a biexponential decay,[58] the fit parameters, $\tau_1$ and $\tau_2$, are shown in the legends of Fig. 4(a) and (b). These transients are recorded at the individual ND luminescence maxima and the dominant decay time (highest intensity in the temporal decay) for each sample is highlighted by bold letters in the legends. Remarkably, for a fixed GaN ND thickness of 4 nm, the decay time decreases by more than two orders of magnitude from $(7.34 \pm 0.10)$ µs for the thickest AlN barriers down to $(40 \pm 7)$ ns for its thinnest counterpart, as predicted for the IFGARD approach.[40] Subsequently, if one decreases the ND thickness (4 – 1 nm) for the samples with the thinnest AlN barrier of 1 nm, the decay time drops further down to approximately 0.7 ns for all three samples. All those samples that are part of the confinement regime feature very similar transients, presented in Fig. 4(b).

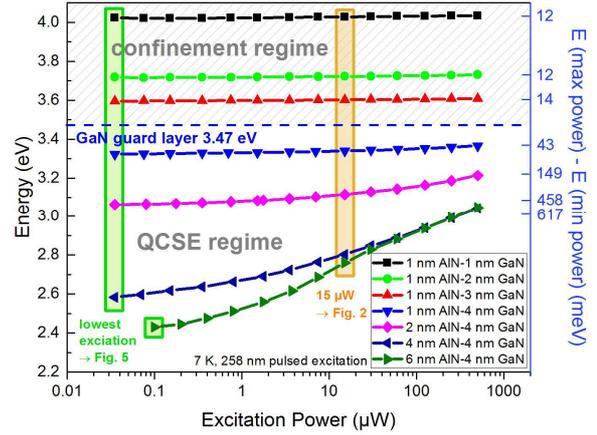

**FIG. 3.** The emission peak positions of the nanodisc (ND) samples are plotted for excitation powers between 0.035 µW and 500 µW. The total emission energy shift between the lowest (min) and highest (max) excitation power is given on the right, vertical axis (blue) representing the achieved polarization field screening by excited charge carriers for the QCSE regime and the effect of ND-band filling for the confinement regime.

## V. DISCUSSION

It is important to note that a 4-nm-thick GaN ND is not able to significantly quantum-confine excitons by its dimensions in absence of the QCSE.[4,49–51] Therefore, all IFGARD samples with a ND thickness of 4 nm would (without the cone-shape of the embedding NWs) emit at a luminescence energy close to the free-exciton-emission energy of bulk GaN (3.48 eV[52,53]). This free exciton emission energy is marked by the luminescence of the bulk-like GaN guard layers in all recorded spectra and represents the transition between the QCSE and the confinement regime. In the confinement regime, the NDs show no indication of built-in fields and the overall energy shift to higher values for thinner NDs is solely caused by the quantum confinement without being affected by a counteracting QCSE red shift. This fact is demonstrated by the power-dependent μ-PL measurements analyzed in Fig. 3. Here, a screening of the internal electric field by excited charge carriers at high excitation powers becomes visible for the QCSE regime, only. The screened QCSE energy drastically reduces for decreasing AlN barrier thicknesses. Within the QCSE regime, the internal electric field strength is a consequence of two effects: first, the unequally sized AlN-GaN interfaces due to the cone-shape of the NWs on either side of each AlN barrier lead to a non-zero net-polarization field inside the NDs, similar to the IFGARD quantum dot case theoretically described by Hönig et al.[40] as the total number of space charges on both interfaces differ from each other. Second, the pronounced cone-shape angle α [compare Fig. 1(b)] in the samples with thicker AlN barriers causes more unequal AlN-GaN interfaces further strengthening the net-polarization as explained in the first case. Hence, the whole QCSE regime features a strong dependence of the ND luminescence energy on the AlN barrier thickness in Fig. 2 and strong screening effects by enhanced excitation powers, cf. Fig. 3. Here, even the highest pump powers (already leading to a moderate sample degeneration under long-time exposure) cannot fully screen the built-in polarization fields in the samples with AlN barrier widths exceeding 1 nm - similar to the unattainable full carrier screening in conventional GaN/AlN QW structures.[31]

Decreasing the ND thickness further reduces the field strength within the NDs [compare Fig. 1(a)] as the finite lateral dimensions of the equally sized "capacitor plates" (due to the finite dimensions of the NWs) become less important. Therefore, the QCSE in the whole ND series becomes negligible resulting in the convergence of the total energy shifts against 12 meV (visible in Fig. 3) indicating a ND-band-filling effect.[59–61] Hence, the overall emission-peak positions of the confinement regime samples in Fig. 2 are purely caused by quantum confinement plus a ND-band filling. This is similar to what was achieved earlier for non-polar GaN/AlN QWs,[29,62–64] which still exhibit lower emission energies than the investigated IFGARD NDs. Besides the aforementioned influence of the decreasing

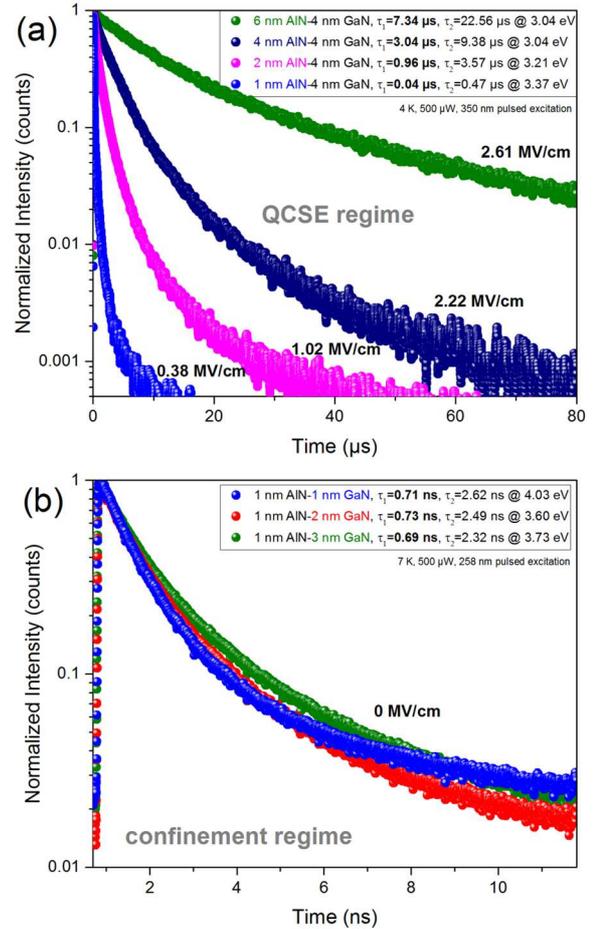

**FIG. 4.** The time-resolved μ-photoluminescence (μ-TRPL) transients of the samples within the QCSE regime shown in (a) reveal the decreasing exciton decay times due to smaller residual electric field strengths in samples with decreased barrier widths. (b) μ-TRPL transients of the samples within the confinement regime show a much smaller, constant decay time due to the eliminated electric field strength. The exciton lifetimes for each biexponential decay are given in the legend; the dominant lifetime is highlighted by bold numbers. All μ-TRPL transients are recorded at the individual ND luminescence maxima.

QCSE on the spectra in Fig. 2, we obtain an increased overlap between the electron and hole occupation probability density, leading to an enhanced exciton-decay rate. In turn, this leads to a strong increase in absolute intensity for the samples with decreased AlN-barrier widths in the QCSE regime. Additionally, the decreased electric dipole moment of the exciton couples less efficiently to defect charge fluctuations in the ND vicinity, which is one mechanism that reduces the FWHM for the emission peaks of the samples related to the QCSE regime.[18] Furthermore, the 1-nm-thick AlN barriers and 4-nm-thick GaN NDs (Fig. 2, blue) exhibit multiple LO-phonon replicas ($E_{LO}$ = 94 meV) whereas all samples within the confinement regime do not show any phonon replicas. While the visibility of multiple LO-phonon replicas implicates a large Huang-Rhys-factor and confirms the existence of built-in excitonic electric dipole moments[65] within the QCSE-regime NDs, their non-visibility for the confinement-regime NDs indicates the absence of a built-in excitonic electric dipole moment.

When the thickness of the NDs in the confinement regime is decreased, the emission energy even exceeds 4 eV due to the strong confinement in absence of the QCSE. The reduction of the absolute intensity (for a constant pump power) can straightforwardly be understood: As the number of NDs (40) is conserved, the photon-absorption volume is reduced. At the same time, the FWHM of the NDs' luminescence peaks within the confinement regime rises. This can be understood considering our quantitative STEM analysis showing random ND-thickness fluctuations of ±1 monolayer thickness that have the strongest impact on the confinement energy of the excitons within the thinnest NDs.[66]

After having separated the QCSE from the confinement effect, the remaining internal electric field strength can be estimated for individual NDs of the QCSE regime simply by the energetic difference between the GaN-guard-layer-luminescence energy and the respective ND emission energy. This difference represents the energy of the electric dipole moment of one exciton in the built-in polarization field. As shown by Nakaoka et al.,[67] the quadratic polarizability of the exciton is only of relevance for electric field strengths up to 100 kV/cm, causing energetic shifts of up to 3 meV. Therefore, we can safely neglect the polarizability of the ND exciton and roughly approximate it by a solid electric dipole moment. Here, we estimate the maximal possible dipole moment $\mu_z$ by the thickness of the NDs times the elemental charge $e$, being $\mu_z \approx 4$ nm · $e$ for the QCSE regime.[68] Hence, the energetic luminescence offset to the 3.47 eV of the bulk-like GaN guard layers $E_{GaN} - E_{ND} = \mu_z F_z$ [69] has to be divided by $\mu_z$ to give an easy approximation of the built-in field strength, that is noted in green in Fig. 5. This approximation should become more reliable for higher field strengths, as the true electric dipole moment of the exciton converges towards the approximated 4 nm · $e$ with higher field strengths. Nevertheless, the estimated 2.61 MV/cm in the IFGARD QW with 6 nm AlN barriers are still more than two-times smaller than the field strengths in comparably thick conventional GaN/AlN QWs.[26,70,71] Although this discrepancy might lead to the assumption of smaller electric dipole moments, i.e., $\mu_z < 2$ nm · $e$ this would be unrealistically small for a 4 nm thick ND, demonstrating the IFGARD effectively reducing the field strength in all samples. Moreover, the overall decrease of the decay time from 7.34 μs down to 0.7 ns [in Fig. 4] correlates with the decreasing electric field strength of the NDs from (a) 2.61 MV/cm over 0.38 MV/cm down to approximately zero for the whole confinement regime (b) with a constant effective decay time of approximately 0.7 ns. Figure 5 summarizes all discussed results: As expected, for a pure confinement effect,[62–64] the luminescence energies (black squares) in the confinement regime converge towards the bulk-like GaN-guard-layer luminescence energy of 3.47 eV.

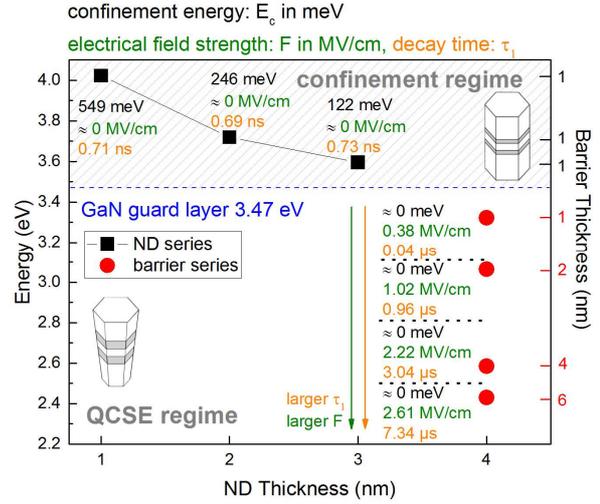

**FIG. 5.** The graphical overview summarizes all results of this work and illustrates the correlation between barrier and nanodisc thicknesses, the nanodisc emission energies (within the confinement regime, marked by black squares and the QCSE regime by red dots), the unscreened electrical field strengths (green values), the exciton confinement energies (black values), and the dominant decay times (orange values). Additionally, the geometric nanowire structure in the respective regimes is indicated.

Finally, it is important to understand that the (111)Si-(0001)GaN interface is expected to build up space charges at the bottom of the NWs. The resulting net-bottom-interface charge causes a field strength that strongly diminishes over the 650 nm distance of the IFGARD-ND stack according to $F \sim \frac{1}{r^3}$ and has therefore only a negligible influence on the entire ND luminescence. This point-charge approximation would not be valid for IFGARD QWs, where the distance to the wafer interface would be small if compared to the wafer diameter. Hence, in such QW samples the bottom hetero-interface should be avoided, e.g., by homo-epitaxial growth or a subsequent laser lift-off process,[72] seeking to reestablish the fundamental symmetry of the IFGARD approach.

## VI. SUMMARY

In summary, we experimentally demonstrate that the IFGARD represents a beneficial method to suppress the internal electric fields in polar heterostructures such as nanodisc (ND) structures. As a result, the QCSE is eliminated, reflected by a shift of the emitted photons to higher energies as compared to conventional quantum wells with matchable thickness and by an enhanced oscillator strength that causes a distinct intensity boost and a decrease in decay times by more than two orders of magnitude for a constant ND thickness. Furthermore, we show that in the presented cone-shaped NW structures, the internal electric field becomes adjustable by the thickness of the AlN barriers. Therefore, the IFGARD opens the possibility to effectively tune the QCSE in polar semiconductor heterostructures without the need for the growth of extraordinary phases or epitaxy on non-polar surfaces.

## ACKNOWLEDGMENTS

We acknowledge support from the Deutsche Forschungsgemeinschaft (DFG) within the Collaborative Research Center 787 (CRC 787). J.M. acknowledges financial support from the JLU Gießen via the graduate scholarship.

## AUTHOR CONTRIBUTIONS

S. S. wrote the manuscript, performed the measurements, and analyzed the presented data. S. S., G. H., and G. C. discussed and interpreted the data. G. H., G. C., and A. H. are the inventors of the IFGARD. J. M., P. H., J. T, J. S., and M. E. developed the NDs in NW sample structure, produced the samples and contributed to the overall data interpretation. T. G. and A. R. supplied the STEM analysis. G. H., G. C., A. H., and M. W. supervise the IFGARD project.